\documentclass[english,english,prstab,twocolumn,superscriptaddress,longbibliography]{revtex4-1}
\usepackage[T1]{fontenc}
\usepackage[latin9]{inputenc}
\usepackage{array}
\usepackage{units}
\usepackage{amsmath}
\usepackage{graphicx}

\makeatletter

\providecommand{\tabularnewline}{\\}

\makeatother

\usepackage{babel}
\begin{document}
\title{Reconstruction of Storage Ring 's Linear Optics with Bayesian Inference}
\author{Yue Hao\thanks{haoy@frib.msu.edu}}
\affiliation{FRIB/NSCL, Michigan State University, East Lansing, MI 48824}
\author{Yongjun Li}
\affiliation{Brookhaven National Laboratory, Upton, NY 11973}
\author{Michael Balcewicz}
\affiliation{FRIB/NSCL, Michigan State University, East Lansing, MI 48824}
\author{L{\'e}o Neufcourt}
\affiliation{FRIB/NSCL, Michigan State University, East Lansing, MI 48824}
\affiliation{Department of Statistics and Probability, Michigan State University,
East Lansing, MI 48824}
\author{Weixing Cheng}
\affiliation{Brookhaven National Laboratory, Upton, NY 11973}
\begin{abstract}
A novel approach of accurately reconstructing the storage ring's linear
optics from turn-by-turn (TbT) data containing measurement error is
introduced. This approach adopts the Bayesian Inference process based
on the Markov Chain Monte-Carlo (MCMC) method, which is widely used
in data-driven discoveries. By assuming a preset accelerator model
with unknown parameters, the inference process yields their posterior
distribution. This approach is demonstrated by inferring the linear
optics Twiss parameters and their measurement uncertainties using
a set of data measured at the National Synchrotron Light Source-II
(NSLS-II) storage ring. Some critical effects, such as the radiation
damping rate, the decoherence due to nonlinearity and chromaticity
can also be included in the model and be inferred directly. One advantage
of the MCMC based Bayesian inference is that it can be performed with
one single set of TbT data to achieve both the inferred results and
their uncertainty before a significant machine drift can happen. The
precise reconstruction of the parameter in the accelerator model with
the uncertainties is crucial prior information to improve machine
performance.
\end{abstract}
\maketitle

\section{Introduction}

In accelerator operations, a significant amount of diagnostic data
is recorded to understand the statistical properties of the bunched
charge-particles. The diagnostic data may be the first order moment
(beam centroid) from beam position monitors (BPMs), second order moment
(beam size) or sampling of the beam distribution from the projection
on the varies types of the profile monitors. These data are the only
clue to tune the control knobs to make the accelerator as the machine
we designed.

Among all tuning tasks, the correction of linear optics is one of
the most important tasks to improve accelerator performance. A more
detailed summary of linear optics measurements has been reviewed in
Ref.\citep{Tomas:2017}. There are various established methods used
to characterize the linear optics experimentally, using the Turn-by-Turn
(TbT) BPM data of a storage ring. These include independent component
analysis (ICA)\citep{ICA}, model-independent analysis (MIA)\citep{MIA},
and orthogonal decomposition analysis (ODA)\citep{Castro:1996} for
retrieving the optics functions tunes, dispersions and chromaticities.
The decoherence of the beam during the TbT data measurements can be
suppressed using pulsed excitation\citep{Malina}. Besides the optics
measurement, The TbT BPM data is also used in the impedance measurement\citep{Carla}.
Another well-developed and widely used method is linear optics from
closed orbit (LOCO) \citep{LOCO}, which heavily depends on the lattice
model.

In this article, an alternative method is demonstrated to retrieve
these machine properties using the Bayesian Inference process from
the accelerator diagnostics data. Bayesian Inference is a powerful
tool to infer unknown parameters $\theta=\left(\theta_{1},\theta_{2},...,\theta_{N}\right)$
of a preset model from a measurement set $M$. In this method, all
to-be-inferred parameters are treated as distribution functions with
initial hypothesis $H$, described by its probability $P(H)$. Since
the initial hypothesis is assumed before observing any measurement
data, it is usually referred to as prior probability. After considering
the measurement set $M$, the probability of the hypothesis is modified
and becomes the posterior probability $P\left(H\mid M\right)$. Using
the Bayes' theorem, we have
\begin{equation}
P\left(H\mid M\right)=\frac{P\left(M\mid H\right)\cdot P\left(H\right)}{P\left(M\right)}\label{eq:bayes' theorem}
\end{equation}
here, $P\left(M\mid H\right)$ is the probability of the observing
$M$ assuming the hypothesis $H$ is valid, also known as the likelihood
function. $P\left(M\right)$ is the marginal probability, which does
not depend on the hypothesis $H$. Alternatively, $P\left(M\right)$
can be interpreted as the normalization factor calculated from the
integral of all possible hypothesis: 
\begin{equation}
P\left(M\right)=\int P\left(M\mid H'\right)dH'.\label{eq:marginal distribution}
\end{equation}

Directly evaluating the Bayes' theorem is difficult, not only because
of the normalization factor in Eq. \ref{eq:marginal distribution}
usually cannot be integrated explicitly, but also due to the possible
high dimension of the unknown parameter space. Therefore, we adopt
the memoryless random step search routine, the Markov Chain Monte
Carlo (MCMC) methods, to sample the posterior probability which is
proportional to 
\begin{equation}
P\left(H\mid M\right)\propto P\left(M\mid H\right)\cdot P\left(H\right)\label{eq:bayes_MCMC}
\end{equation}
without evaluating $P\left(M\right)$. The detailed algorithm of MCMC
is introduced in Appendix A.

We will demonstrate this Bayesian Inference process by determining
the linear optics using the BPMs' TbT data, together with the nonlinear
decoherence and the radiation damping effects. The structure of this
article is outlined below: we first detail the betatron model and
its parameters in the next chapter (chapter II), then illustrate the
inference results of the optics function in chapter III, and discuss
the model selection criteria for the Bayesian Inference in chapter
IV.

The result of the Bayesian Inference, which is the sample from the
posterior distribution of the parameter in the model, can play an
important role in refining the accelerator settings. In the optics
inference example, the posterior distribution of the Twiss parameters
will be essential in understanding how well the optics correction
can achieve based on the measurement. Recently, a Bayesian approach
\citep{Bayesian_linear_optics} was applied for linear optics correction
in a storage ring, based on a set of measured linear optics distortions
with some uncertainties, and it was noticed that a prior distribution
of quadrupole error can be used to specify an optimal regularization
coefficient to prevent overfitting.

\section{Model and Measurement Data}

Consider $N$ BPMs distributed around a storage ring. These BPMs can
record some selected bunches' \citep{Li:2017} TbT trajectory for
$T$ turns after the beam is excited to perform a betatron oscillation.
We can construct a betatron oscillation model to represent the measurement
data without detailed knowledge of accelerators, such as the layout
of the lattice and the magnet strength. In general the Betatron motion
recorded by the $i^{\text{{th}}}$ BPM at $j^{\text{{th}}}$ turn
can be written as:

\begin{equation}
x_{i}(j)=x_{c,i}+A_{i}(j)x_{i}(0)\cos\left(2\pi\nu_{x}j+\phi_{i}\right)+\xi_{i}(j)\label{eq:xj_with_error}
\end{equation}
 Here, $x_{c,i}$ is the closed orbit at $i^{\text{{th}}}$ BPM, $\nu_{x}$
is the horizontal betatron tune and $\phi_{i}$ is the phase constant
at $i^{\text{{th}}}$ BPM. $A(j)$ is introduced to represent the
envelope evolution due to decoherence and radiation damping effect,
which will be addressed later. At each turn, the BPM reading contains
a random error which follows the Gaussian distribution with a zero
mean value and standard deviation $\sigma_{\xi}$.

We treat the first two terms in Eq. \ref{eq:xj_with_error} as our
``model'':
\begin{equation}
x_{i,model}(j)=x_{c,i}+A_{i}(j)x_{i}(0)\cos\left(2\pi\nu_{x}j+\phi_{i}\right)\label{eq:model}
\end{equation}
Then the difference of the measurement data and the model follows
a random Gaussian distribution:
\[
x_{i,data}-x_{i,model}\propto N(0,\sigma_{xi})
\]
for all $j^{\text{th}}$ measurements. The Gaussian random distribution
is centered at zero because the bias of the BPM reading and the beam
closed orbit $x_{c,i}$ cannot be distinguished by the model in Eq.
\ref{eq:model}.

The betatron oscillation part of the model can be re-written in terms
of the Twiss parameters:\begin{widetext}
\[
\left(\begin{array}{c}
x_{\beta,i}\\
x_{\beta,i}'
\end{array}\right)_{j}=A_{i}(j)\left(\begin{array}{cc}
\cos2\pi\nu_{x}j+\alpha_{i}\sin2\pi\nu_{x} & \beta_{i}\sin2\pi\nu_{x}j\\
\left(-\left(1+\alpha_{i}^{2}\right)\sin2\pi\nu_{x}j\right)/\beta_{i} & \cos2\pi\nu_{x}j-\alpha_{i}\sin2\pi\nu_{x}j
\end{array}\right)\left(\begin{array}{c}
x_{\beta,i}\\
x_{\beta,i}'
\end{array}\right)_{0}
\]
\end{widetext}In each turn $j$, the action $J_{i}$ at $i^{\text{th}}$
BPM has the form:

\[
J_{i}(j)=\frac{1}{\beta_{i}}\left(x_{\beta,i}^{2}\left(j\right)+\left(\beta_{i}x_{\beta,i}'\left(j\right)+\alpha_{i}x_{\beta,i}\left(j\right)\right)^{2}\right)
\]

In this form, only information of $x_{\beta}$ can be retrieved from
BPM data, while $\beta$, $\alpha$ and $x_{\beta}'$ are unknown.
We can only infer the combination of these three unknowns. Following
the usual accelerator physics notation, we will only infer the conjugate
variable pair $\left(x_{\beta},P_{x}\right)$, where the conjugate
momentum $P_{x}$ is:
\[
P_{x}\equiv\beta x_{\beta}'+\alpha x_{\beta}
\]
then we can get 
\begin{equation}
J_{i}(j)\beta_{i}=x_{\beta,i}^{2}\left(j\right)+P_{x,i}^{2}\left(j\right)\label{eq:beta_j}
\end{equation}
Using the measurement data at each turn, we can infer the betatron
oscillation amplitude and the conjugate momentum, hence determine
the beta function at each BPM $i$, scaled by the action of the betatron
oscillation. In linear optics approximation, the action $J$ is a
constant for all BPM locations for any turns after scaled by the envelope
factor $A_{i}^{2}(j)$. However, in the experiments at NSLS II, the
centroid of the electron beam is kicked to a fraction of 1 mm, thus
the nonlinearity of the sextupole cannot be neglected. The simulation
shows that the resulting fluctuation of the centroid's action is a
few percent across the BPMs in one turn. We may overcome this difficulty
by average the actions over the number of turns of the measurement
data. The same simulation shows that the action fluctuation reduces
to $6\times10^{-4}$ after average over 100 turns, and $3\times10^{-4}$
over 500 turns. Since the accuracy of the beta function usually is
expected to be $\sim$1\%, we can ignore the fluctuation of the action
after averaging it over the measurement time ($\sim$1000 turns).

Finally, a proper form of $A\left(j\right)$ has to be selected to
reflect obvious damping behavior. We choose to express the factor
inside the exponential function as a polynomial of the turn number:

\begin{equation}
A(j)=\exp\left(-\epsilon_{SR}j-\epsilon_{NL}^{2}j^{2}\right)\label{eq:Damping factor form}
\end{equation}
Clearly, we can link the coefficients $\epsilon_{SR}$ and $\epsilon_{NL}$
to the synchrotron radiation damping and nonlinear decoherence effect
\citep{Meller_decoherence} respectively. It is worthwhile to note
that we took the simplified nonlinear decoherence form (Eq. 18) of
reference \citep{Meller_decoherence}. The justification is included
in Appendix B.

In this simplified model, we only include the transverse motion in
one direction and excluded the effect of chromaticity. The treatment
of synchrotron-transverse coupling will be discussed at the end of
this article.

\section*{Bayesian Inference}

The unknown parameters in the model (as in Eq. \ref{eq:model}) are
noted as $\theta=\left(x_{0,}x_{c},P_{x},\nu_{x},\epsilon_{SR},\epsilon_{NL}\right)$,
along with the standard deviation of the Gaussian distribution $\sigma_{\xi}$.
These parameters will be inferred from the TbT data $X_{data}=\left(x_{0,}x_{1,}\cdots,x_{T-1}\right)$.
Here, the data for each BPM in one measurement is $\sim2000$ turns.
Applying the Bayesian theorem (Eq. \ref{eq:bayes_MCMC}), we have:
\[
P\left(\theta\mid X_{data}\right)\propto P\left(X_{data}\mid\theta\right)P\left(\theta\right)
\]
which can be calculated using the MCMC method.

\begin{figure}
\includegraphics[width=1\columnwidth]{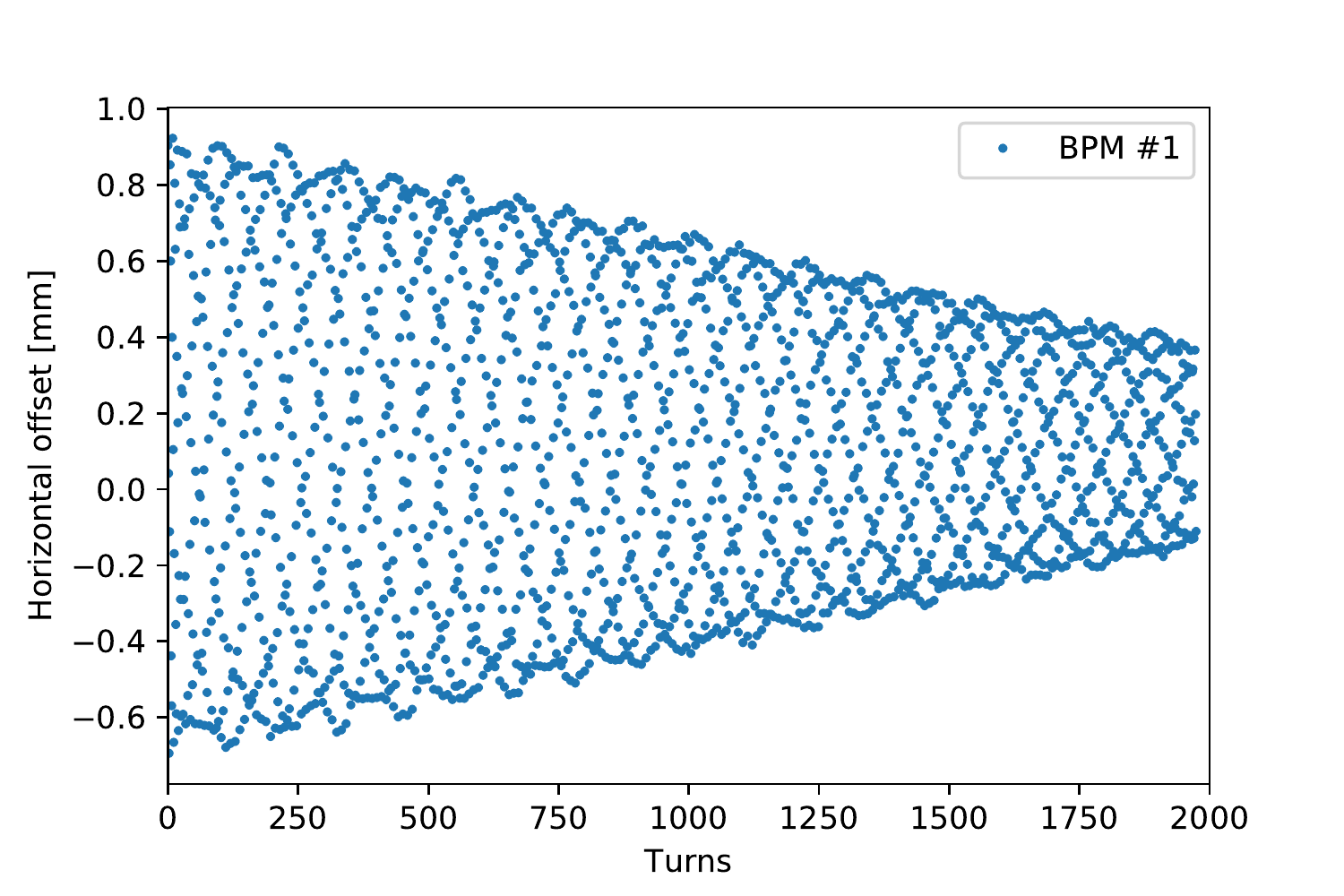}

\caption{BPM reading of the first BPM\label{fig:bpmsignal}}
\end{figure}

We use a set of measured NSLS-II TbT data to demonstrate this method.
The NSLS-II is a 3rd generation light source, which has 30 double
bend achromat cells \citep{NSLS-II:2013}. It is equipped 180 quasi-uniformly
distributed BPMs for the purpose of orbit and lattice monitoring and
control. The beam can be excited with a horizontal and a vertical
fast pulse magnet to perform a free betatron oscillation. The BPMs
are configured with TbT resolution and having a gated functionality
to lock on a selected diagnostic bunch train \citep{Li:2017}. Figure
\ref{fig:bpmsignal} shows one snapshot of TbT data from one of these
180 BPMs.

\begin{figure*}
\includegraphics[width=0.33\textwidth]{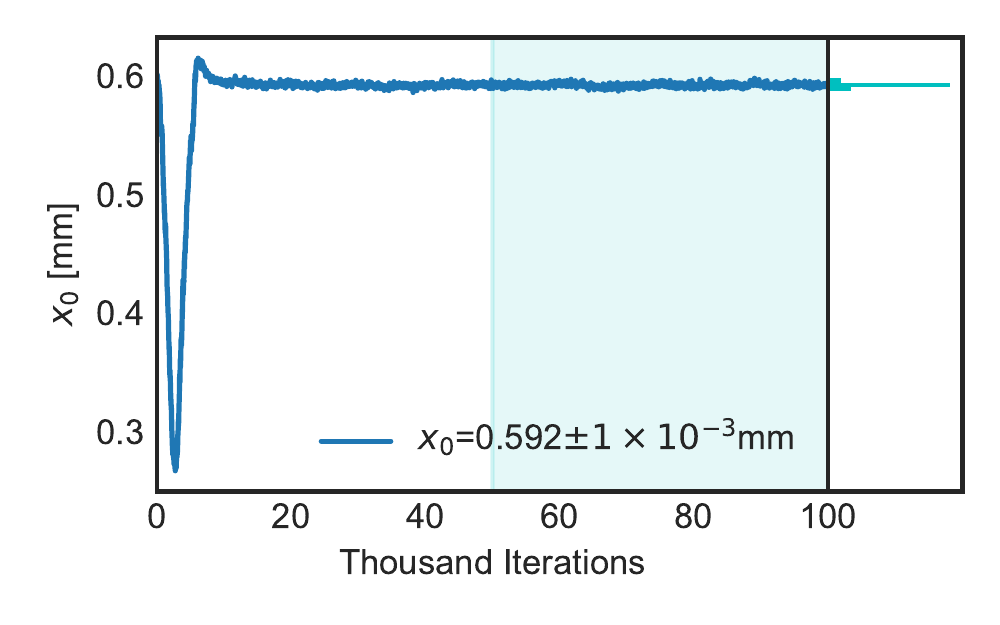}\includegraphics[width=0.33\textwidth]{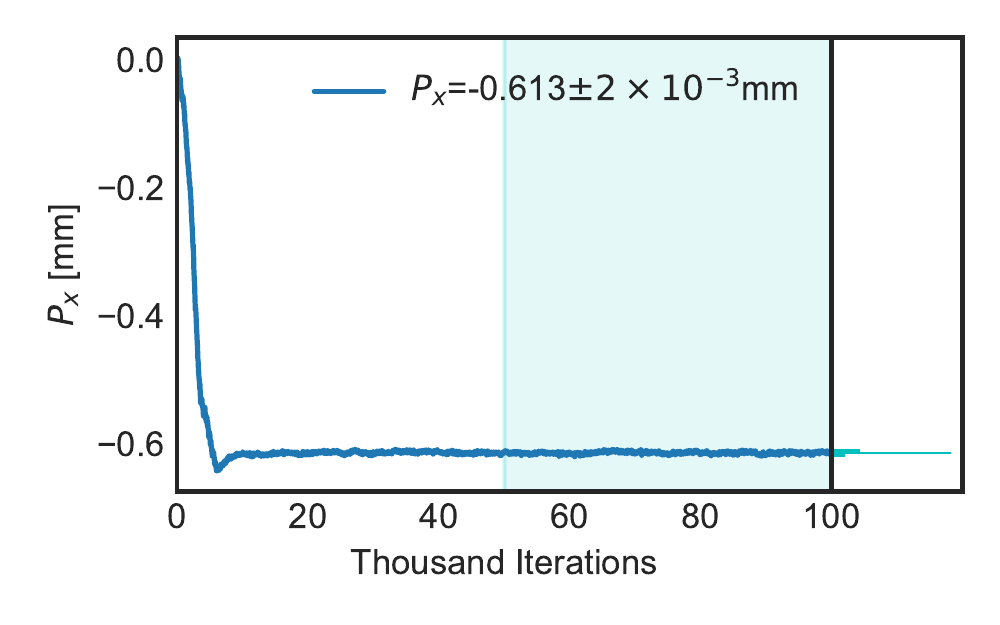}\includegraphics[width=0.33\textwidth]{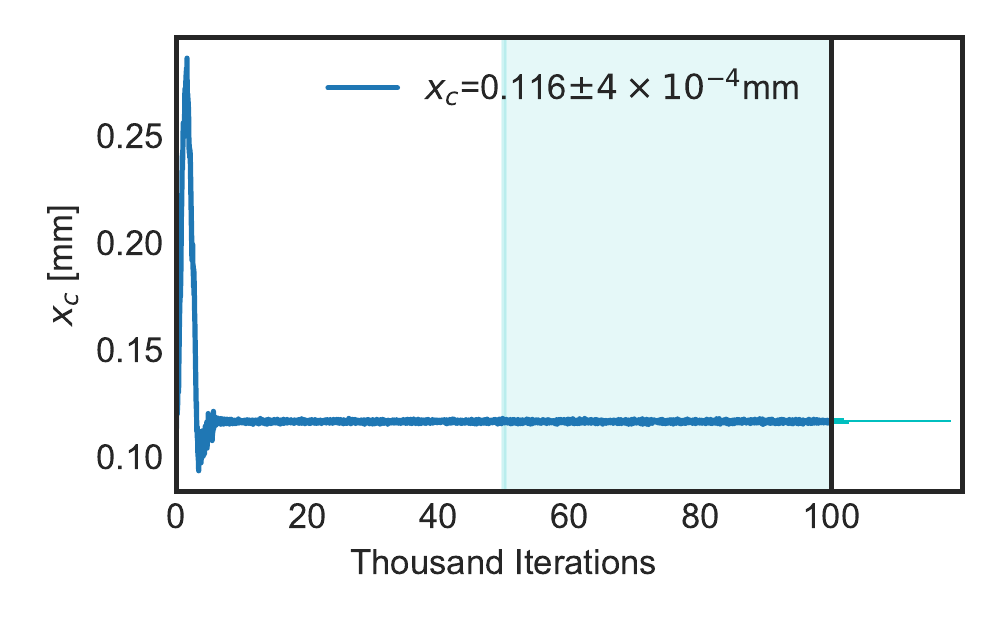}

\includegraphics[width=0.33\textwidth]{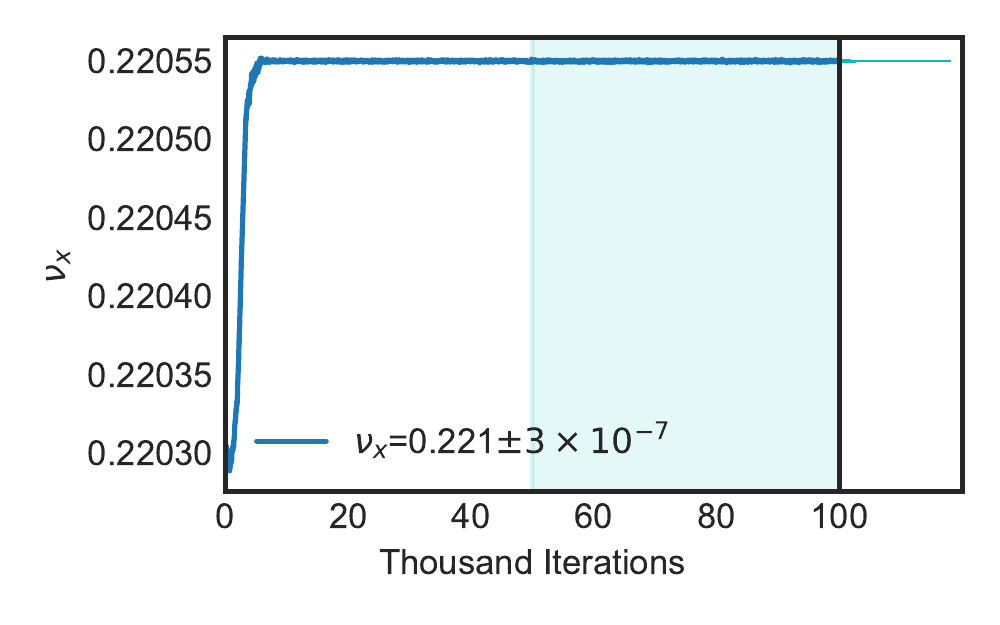}\includegraphics[width=0.33\textwidth]{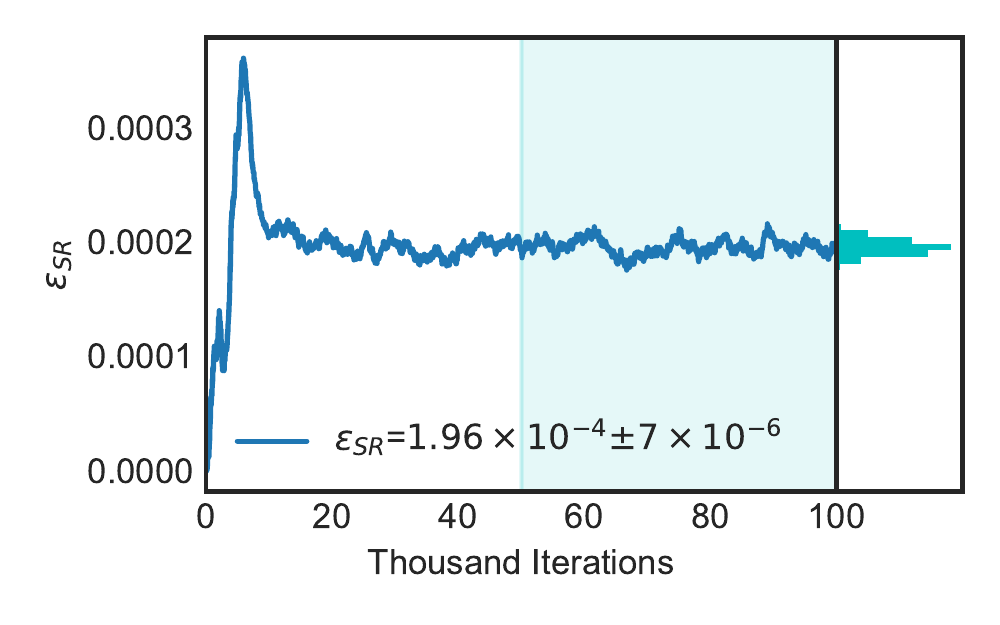}\includegraphics[width=0.33\textwidth]{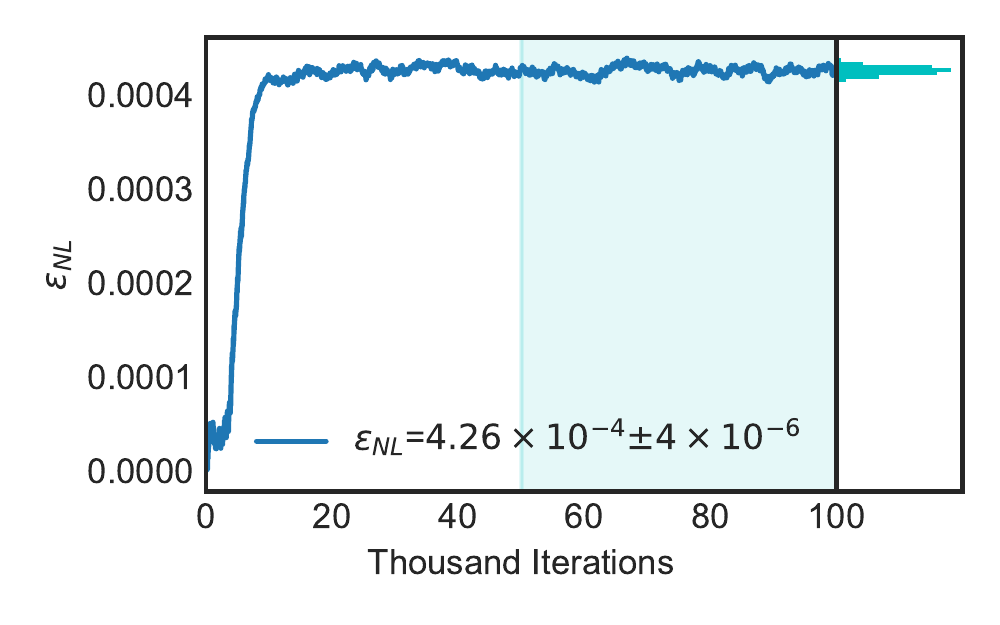}

\caption{MCMC iterations of parameter $\theta$ using the measurement data
of first BPM. From left to right, the top figures are the iterations
for $x_{0},$ $P_{x}$ and $x_{c}$; the bottom figures are for $\nu_{x}$,
$\epsilon_{SR}$ and $\epsilon_{NL}$ respectively. A histogram plot
is attached to the right of the each iteration plot. The histogram
is taken from the second half of the total iteration, which is highlighted
by the shaded area.\label{fig:iterations}}
\end{figure*}

\begin{table}
\caption{Initial values, step sizes and the inference results of parameters
in the model Eq.\ref{eq:model} \label{tab:parameter-values}}

\begin{tabular}{|c|c|c|c|}
\hline 
 & Initial values & Step sizes & Inference results\tabularnewline
\hline 
\hline 
$x_{0}$ (mm) & $X_{data,0}$ & $1\times10^{-3}$ & 0.592$\pm1.4\times10^{-3}$\tabularnewline
\hline 
$x_{c}$ (mm) & $\bar{X}_{data}$ & $1\times10^{-3}$ & 0.116$\pm3.7\times10^{-4}$\tabularnewline
\hline 
$P_{x}$ (mm) & 0 & $1\times10^{-3}$ & -0.613$\pm1.5\times10^{-3}$\tabularnewline
\hline 
$\nu_{x}$ & Peak of DFT & $1\times10^{-7}$ & 0.22055$\pm2.7\times10^{-7}$\tabularnewline
\hline 
$\epsilon_{SR}$ & 0 & $1\times10^{-6}$ & $1.96\times10^{-4}$$\pm7\times10^{-6}$\tabularnewline
\hline 
$\epsilon_{NL}$ & 0 & $1\times10^{-6}$ & $4.26\times10^{-4}$$\pm4\times10^{-6}$\tabularnewline
\hline 
$\delta_{\xi}$ (mm) & $1\times10^{-3}$ & $1\times10^{-3}$ & 0.0164$\pm2.7\times10^{-4}$\tabularnewline
\hline 
\end{tabular}
\end{table}

\begin{figure}
\includegraphics[width=1\columnwidth]{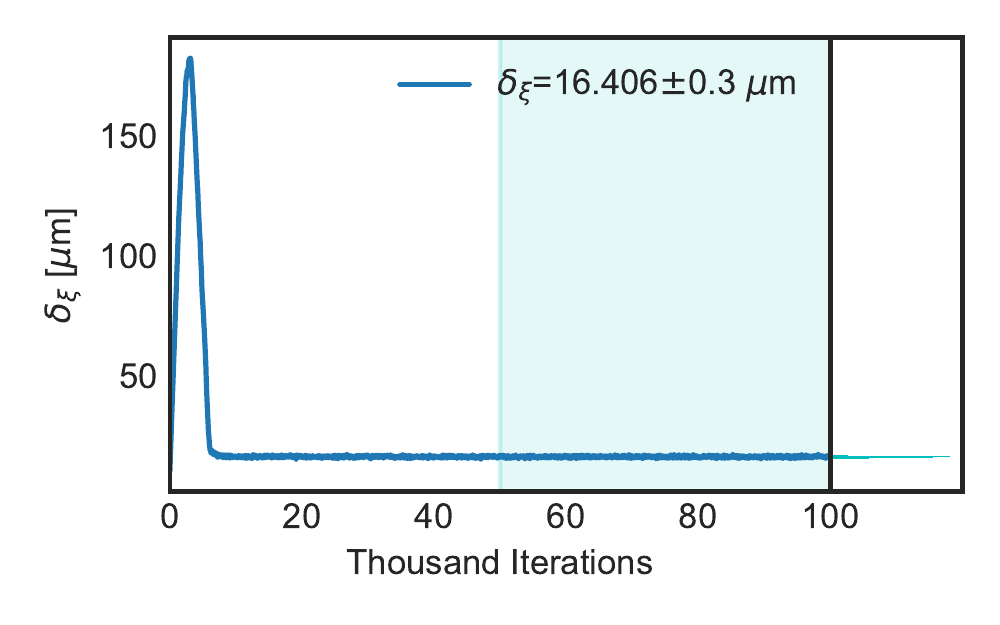}

\caption{MCMC iteration of the width of the Gaussian error.\label{fig:iteration_sigma}}
\end{figure}

The BPM reading reflects the orbit excursion, due to closed orbit,
initial beam kick, the machine properties, and the TbT reading errors.
The BPM random error of one turn is independent of the measurement
of other turns and assumed to be Gaussian random distribution. Therefore
the likelihood $P\left(X_{data}\mid\theta\right)$ can be expressed
as:

\[
P\left(X_{data}\mid\theta\right)\propto\exp\left(-\frac{1}{2\sigma_{\xi}^{2}}\sum_{j=0}^{T-1}\left(x_{data,j}-x_{model,j}\right)^{2}\right)
\]

The prior probability $P\left(\theta\right)$ reflects our foreknowledge
of these parameters before the measurement. For example, it is expected
that the parameter $x_{0}$ should be very close to the first value
of measurement data $x_{data,0}$; the closed orbit parameter should
be close to the average value of the measurement $\bar{X}_{data}$,
while the transverse tune $v_{x}$ is expected to be close to the
discrete Fourier transform (DFT) of the BPM data. In addition, the
prior distribution can also be used to confine the range of the parameter,
such as confining the parameter $\epsilon_{SR}$ and $\epsilon_{NL}$
to be positive. A proper prior will improve the MCMC convergence efficiency
and conversely, while incorrect prior will bias the posterior distribution.

In this model, we do not imply any specific prior distribution other
than requiring $\epsilon_{SR}$ and $\epsilon_{NL}$ to be positive.
The initial value and the step size of each parameter are given in
Table \ref{tab:parameter-values}. The initial value of each parameter
reflects the 'best-guess' value for the corresponding parameter. The
choice of step size also plays an important role in the inference
process. A small step size would require long iteration steps and
long computation time, while a large step size will result in failing
to reach convergence. A proper choice of the step size of one parameter
should be smaller than the order of the uncertainty of the parameter.
In this example, the step sizes of $x_{0}$, $x_{c}$,$P_{x}$ $\delta_{\xi}$
are set to 1 micron, based on a reasonable guess that the NSLS-II
BPMs have TbT resolution around 10 microns. The proper step sizes
of other parameters are determined by various trials with simulations.

Figure \ref{fig:iterations} and Figure \ref{fig:iteration_sigma}
show the MCMC iterations for the data from one of the BPM. All figures
illustrate convergence results after 25000 iterations. The parameter
values of each iteration after convergence reaches (cyan shaded area
in each figure) can be used to calculate the posterior distribution.
The histogram can be plotted to represent the distribution and is
attached to the right of each iteration plot. The converged results
and their standard deviations are listed in the last column of Table
\ref{tab:parameter-values}.

It is important to note that the uncertainties are obtained from one
snapshot of data ($\sim2000$ turns) using the Bayesian inference.
Comparing with other methods, such as ICA, MIA, and ODA, each snapshot
can give one measurement result and the uncertainty is obtained from
repetitive measurements. However, the machine has been observed continuously
drifting during repetitive data collection. Therefore the uncertainty
from repetitive measurement is over-estimated with the traditional
methods. The overestimated uncertainty could potentially affect the
ultimate performance of linear optics correction \citep{Bayesian_linear_optics}.

A notable feature of the inferred tune can be found in the bottom
left plot of Figure \ref{fig:iterations} and the inference results
in Table \ref{tab:parameter-values}. The initial guess of the transverse
tune from the peak DFT position of the BPM's data is not precise enough
due to a limited length of the data. The inferred tune has much higher
accuracy ($\sim10^{-7}$ compared with $\sim1/T$, $T$ is the number
of turns). The high accuracy of the inferred tune using Bayesian Inference
is because of fitting the data with the given model, which is similar
to the numerical analysis of fundamental frequency (NAFF) \citep{NAFF}
method. The NAFF method aims at precisely searching for a series of
orthogonal frequencies using one snapshot of the measurement data.
The uncertainty can be retrieved from multiple measurements \citep{langner,Biancacci,Persson}.
With the MCMC algorithm, Bayesian Inference achieves not only a precise
tune but a sound quantification of the uncertainty with one single
snapshot. In addition to the pure betatron oscillation, it is convenient
to achieve additional envelop profiles according to the physics knowledges,
which is represented by the envelope factor $A\left(j\right)$ in
our model. It is worthwhile to note that many forms of envelope function
will change the frequency of the oscillation, although the change
may be small enough to neglect in real applications.

\begin{figure}
\includegraphics[width=1\columnwidth]{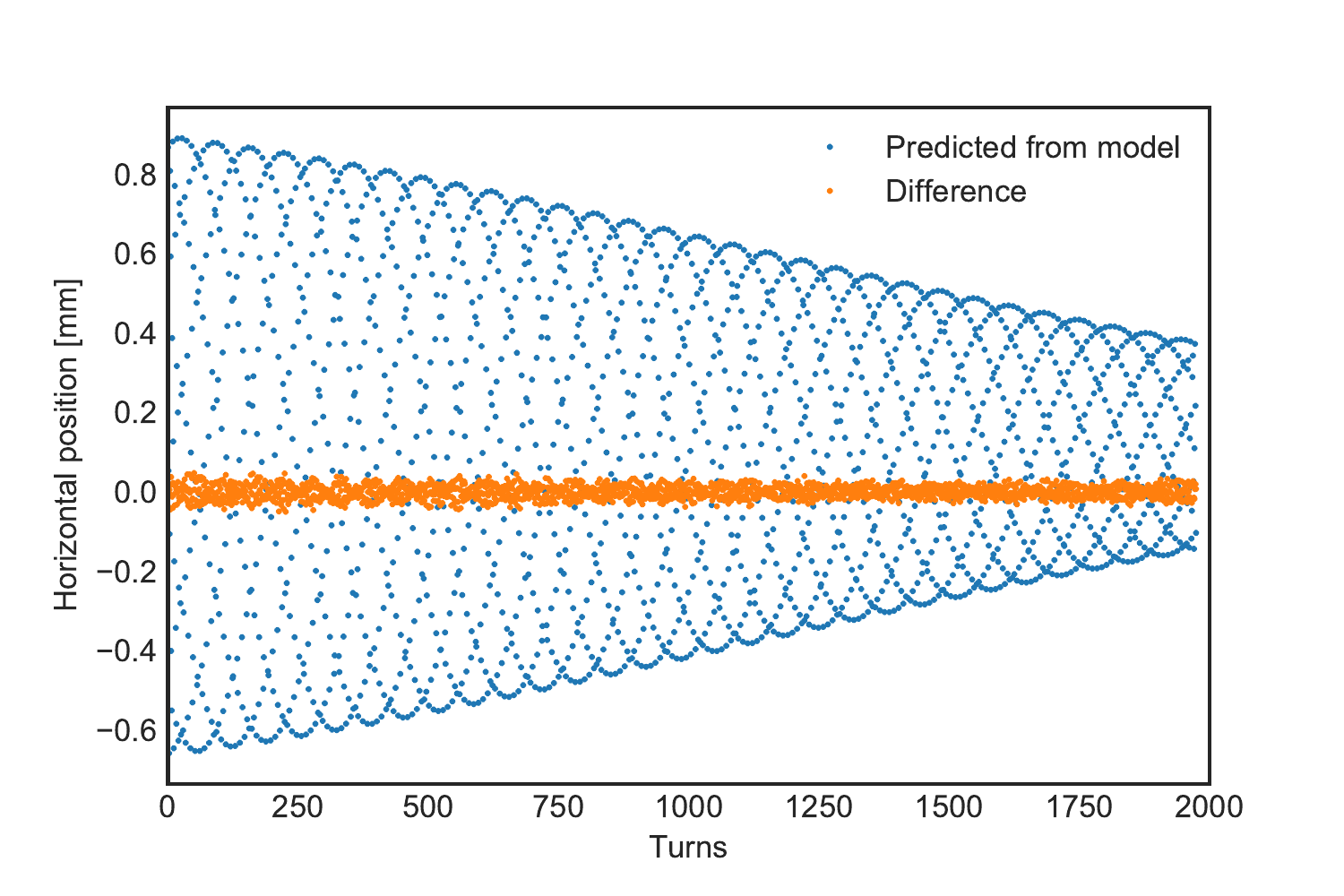}

\caption{Reconstruction of the first BPM reading using model with inferred
parameter (blue) and the difference with the measurement data (orange)\label{fig:difference}}
\end{figure}

\begin{figure}
\includegraphics[width=1\columnwidth]{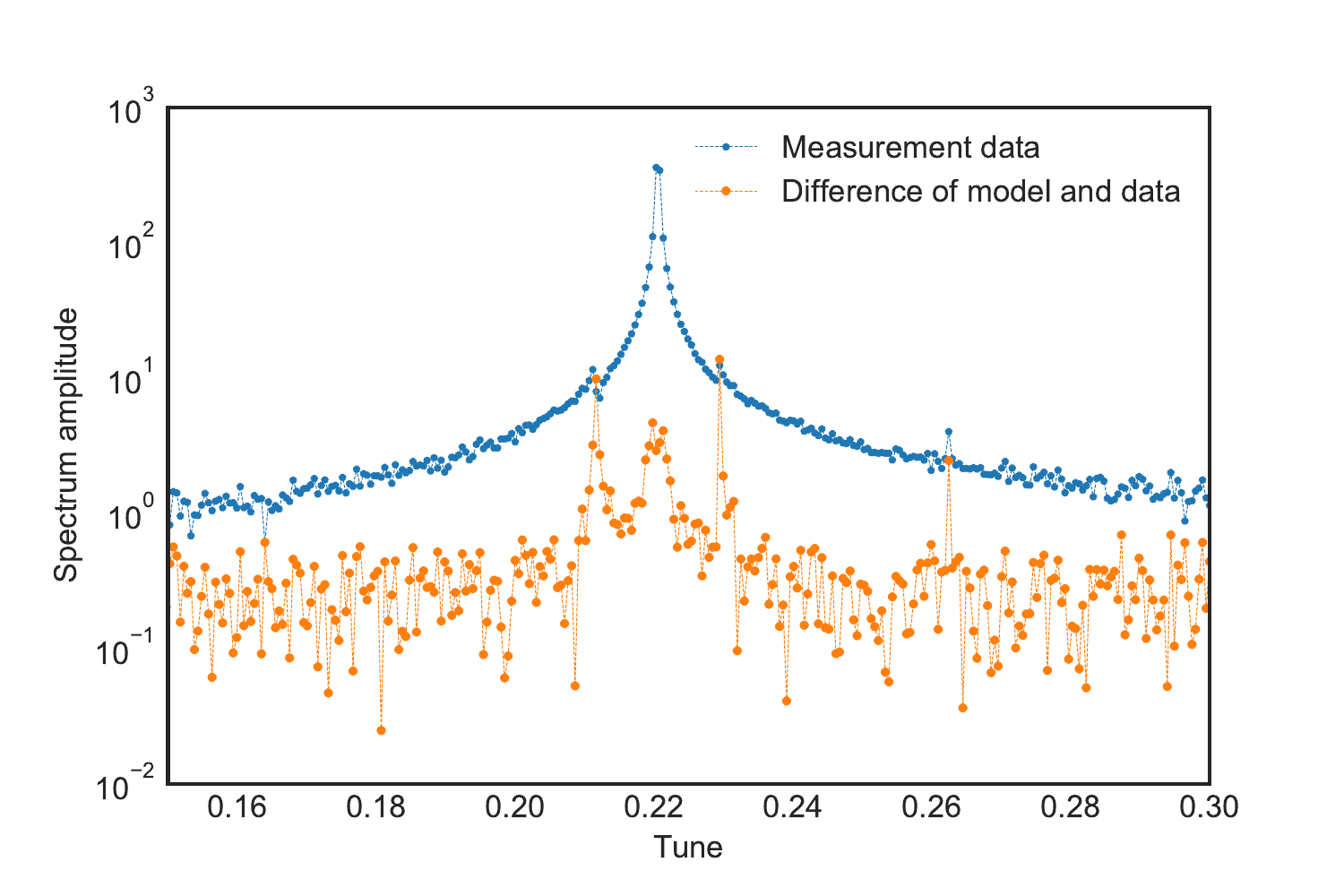}

\caption{The frequency spectrum of measurement data of the BPM (blue points)
and the difference between measurement and model with inferred parameters
(orange points)\label{fig:difference-dft}}
\end{figure}

Figure \ref{fig:difference} shows that the model with the inferred
parameter represents the measurement data plausibly, with the difference
plotted in orange dots. The difference is assumed as a Gaussian random
number and is characterized by its standard deviation $\sigma_{\xi}$.
However, our model in Eq. \ref{eq:model} cannot reflect all physics
in the accelerator completely. The difference in the measurement data
and the model prediction must contain other signals rather than pure
BPM noises. Therefore, the estimation of $\sigma_{\xi}$ only gives
an upper bound on the actual BPM noises. Fourier analysis gives the
frequency content of the difference data as shown in figure \ref{fig:difference-dft}.
Clearly, the transverse coupling signal at a tune of about 0.26 is
observed, as well as two synchrotron sidebands, about 0.009 away from
the transverse tune 0.222.

As pointed out in the previous section, the initial position $x_{0}$,
and its conjugate momentum $P_{x,0}$ can be inferred from the data.
Then, as shown in Eq. \ref{eq:beta_j}, the product of the beta function
and the action can be determined for this BPM:

\begin{align*}
\beta J & =\left(x_{0}-x_{c}\right)^{2}+P_{x}^{2}\\
 & =0.602\pm0.003\text{mm}^{2}
\end{align*}

\begin{figure}
\includegraphics[width=1.1\columnwidth]{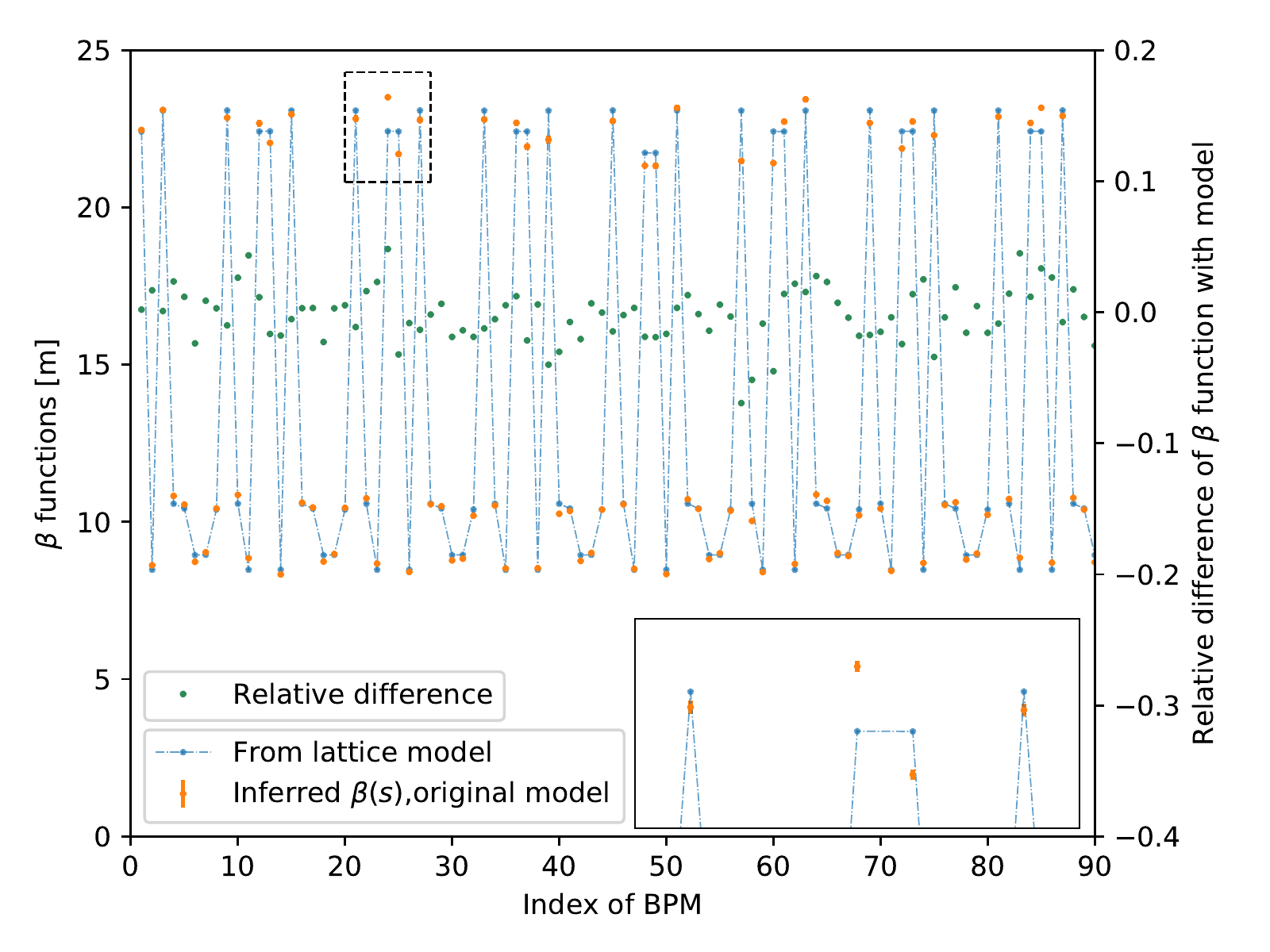}

\caption{The inferred beta function at first 90 BPMs and its standard deviation
error (orange dots), with the beta function calculated from bare lattice
(in blue doted lines) as a reference. The relative difference of the
inferred beta function and the model is shown in green dots.\label{fig:inferred beta function}}
\end{figure}

The inference process is repeated and get the $\beta J$ for all BPMs.
The action of the centroid can be averaged over the measurement period.
The fluctuation of the average value is less than $4\times10^{-4}$
for all BPMs, according to the particle simulation using the NSLS-II
lattice model. The action is almost constant when compared with the
uncertainty of the $\beta J$. However, we can not retrieve the value
of $J$ from the BPM data. We have to choose a constant $J$ so that
the average of the inverse beta function $\overline{1/\beta}$ equals
that calculated from the accelerator model. This step of scaling only
aims at producing inferred beta function in the familiar range. Figure
\ref{fig:inferred beta function} shows the inferred beta functions
and its error bars, which is barely visible in this plotting scale.
For better visibility, only beta functions at the first 90 BPMs are
plotted. The average standard deviation of beta functions is about
0.47\%, while the difference of beta function from the model and the
measurement is 10\% peak to peak, as shown in the green dots of Figure
\ref{fig:inferred beta function}. Therefore there is a large potential
for the optics to be corrected. In the optics correction process,
the uncertainty of the optics function and phase advance, retrieved
from the Bayesian Inference, can be used directly to define the regularization
coefficients in order to prevent overfitting issue, as pointed out
in \citep{Bayesian_linear_optics}.

\begin{figure}
\includegraphics[width=1\columnwidth]{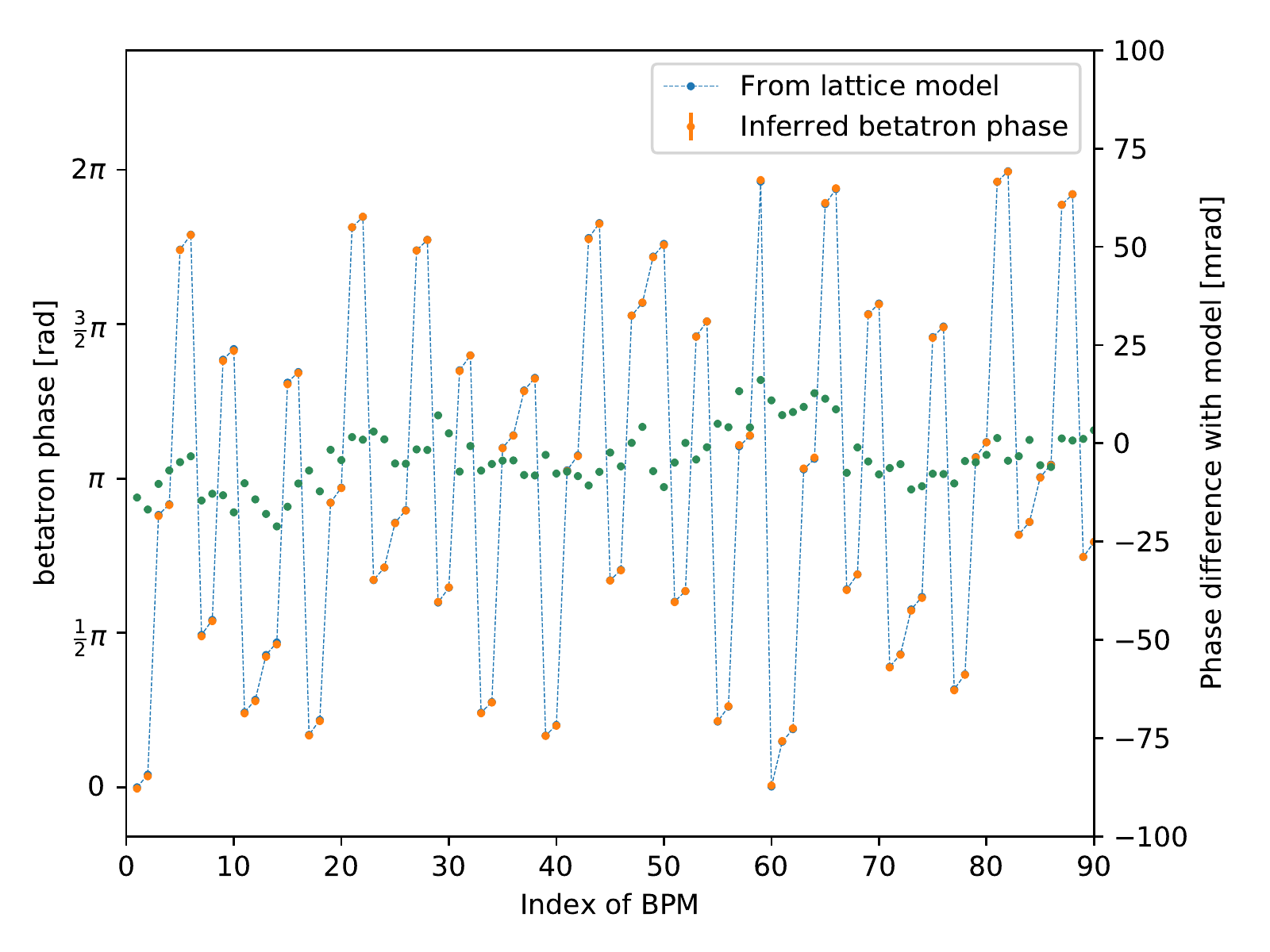}

\caption{The inferred betatron phase at first 90 BPMs (orange dots), with the
betatron phase advance calculated from bare lattice (blue doted line)
as reference. The phase difference is shown in the green dots.\label{fig:inferred betatron phase}}
\end{figure}

The betatron phase $\phi_{\beta}$ is calculated as:
\[
\phi_{\beta}=\arctan\left(\frac{P_{x}}{x_{0}-x_{c}}\right)
\]
Figure \ref{fig:inferred betatron phase} compares the inferred betatron
phase and the phase calculated from the accelerator model. The standard
deviations of the betatron phase vary from $1.5\times10^{-3}$ and
$3\times10^{-3}$ rad, therefore they are not visible in the figure.

\begin{figure*}
\includegraphics[width=0.33\textwidth]{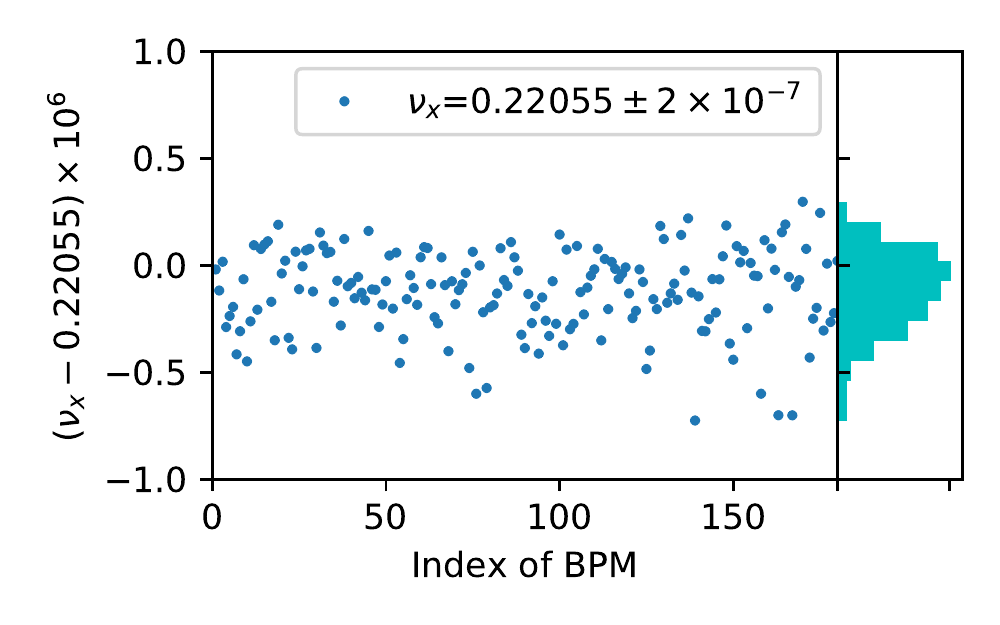}\includegraphics[width=0.33\textwidth]{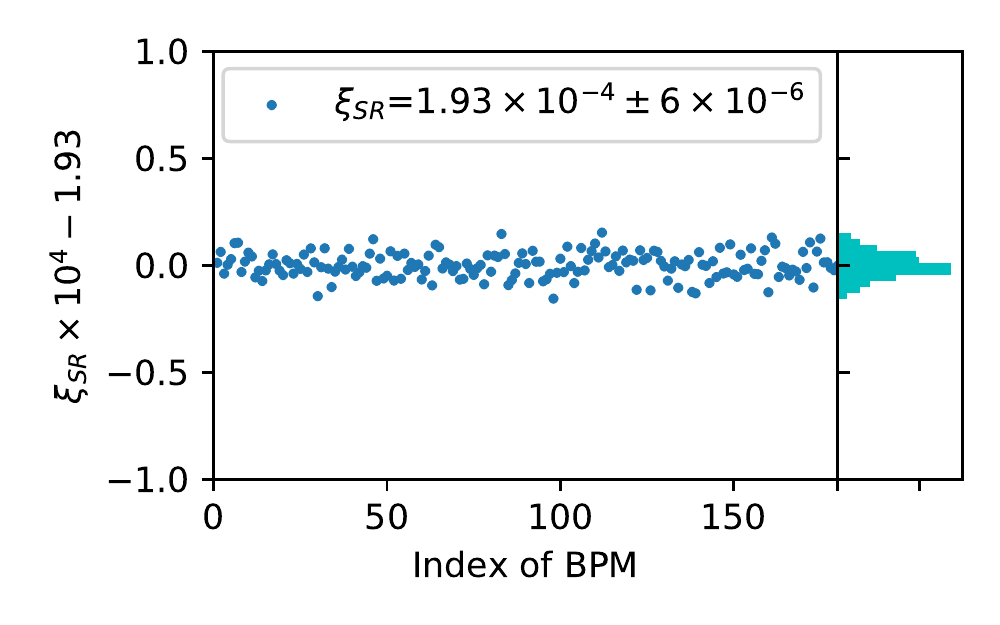}\includegraphics[width=0.33\textwidth]{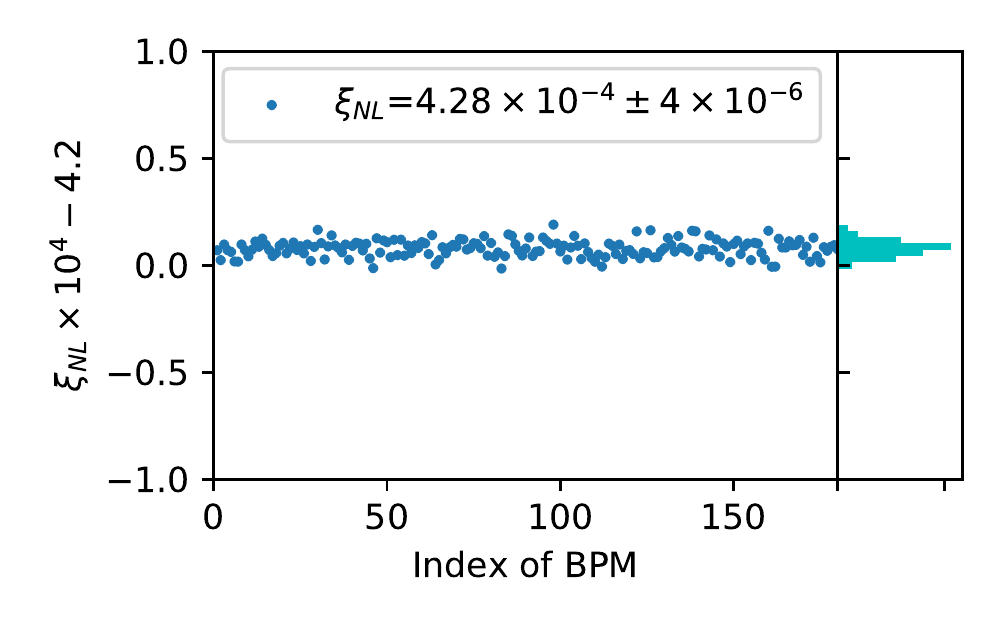}

\caption{The tune, $\epsilon_{SR}$ and $\epsilon_{NL}$ inferred from all
180 BPMs. For easier reading, they are rescaled and shifted to the
range {[}-1,1{]}. The quantity bing plotted are left: $\left(\nu_{x}-0.22055\right)\times10^{6}$,
middle: $\epsilon_{SR}\times10^{4}-1.93$ and right: $\epsilon_{NL}\times10^{4}-4.20$.\label{fig:tune_epsilon_all_bpm}}
\end{figure*}

\begin{table}
\caption{Comparison of standard deviation of the mean values of 180 inferences
and the average of the standard deviation in each inference. \label{tab:inference_individual_all}}

\begin{tabular}{|c|>{\centering}m{0.4\columnwidth}|>{\centering}m{0.4\columnwidth}|}
\hline 
 & Standard deviation of mean value of 180 inferences & Average of the uncertainty of each mean value\tabularnewline
\hline 
\hline 
$\nu_{x}$ & $1.9\times10^{-7}$ & $3.4\times10^{-7}$\tabularnewline
\hline 
$\epsilon_{SR}$ & $5.5\times10^{-6}$ & $6.0\times10^{-6}$\tabularnewline
\hline 
$\epsilon_{NL}$ & $3.6\times10^{-6}$ & $4.1\times10^{-6}$\tabularnewline
\hline 
\end{tabular}
\end{table}

We expect a priori that the parameter $\nu_{x}$, $\epsilon_{SR}$,
and $\epsilon_{NL}$ should be common to the entire ring, as they
are the 'integrating' parameters and reflects the dynamics of the
entire ring. Thus it is a useful step to check that the inference
produces consistent values across all BPMS. Figure \ref{fig:tune_epsilon_all_bpm}
shows the inferred mean values of $\nu_{x}$, $\epsilon_{SR}$, and
$\epsilon_{NL}$ from data of 180 BPMs, and their histograms are attached
to the right. Table \ref{tab:inference_individual_all} illustrates
that the deviation of the mean values of the 180 inference results
is less than the average of the uncertainty (standard deviation) of
each mean value. It supports our belief that we can infer very similar
values of $\nu_{x}$, $\epsilon_{SR}$, and $\epsilon_{NL}$ from
all BPMs, as it should be from the physics behind the model.

\section*{Model Selection}

The results in the previous sections are based on the assumed model
in Eq. \ref{eq:model}. A valid model is key to the success of Bayesian
inference. The model itself can be also viewed as the strongest prior
information, which enhances a finer treatment than considering the
accelerator as a black box. The confidence of placing these models
based on physics knowledge, instead of other general models such as
an artificial neural network, relies on the belief that the accelerator
model can represent most of the physics in an accelerator. In this
chapter, we discuss the strategy of model selection in Bayesian Inference.

One could choose a very precise model with detailed accelerator elements
such as the magnets, cavities and the drift space between them. The
currents, voltage and the geometric information (length, distances)
will be the unknown parameters to be inferred from measurement data.
However, it is impractical, at least for the computation power nowadays,
to infer such a high-dimensional problem, as there are usually thousands
of knobs to describe accelerator lattice to be inferred, even with
strong prior assumptions and a large amount of data. Meanwhile, in
most cases, the goal of optimizing accelerator operation is not to
know every detail of accelerators, but the key parameters that affect
the performance, for instance, the beam behavior at the interaction
point in colliders or at the insertion devices of synchrotron light
sources.

Therefore it is reasonable to choose the model to represent the dynamics
of interest with few parameters. In our example the previous chapter,
the focus is only on betatron motion and its damping envelope, as
in Eq. \ref{eq:model}. In many existing methods, the damping envelope
is ignored, which corresponds to adjust our model by forcing $A\left(j\right)\equiv1$.
This reduces the model to 
\begin{equation}
x_{i,model}(j)=x_{c,i}+x_{i}(0)\cos\left(2\pi\nu_{x}j+\phi_{i}\right)\label{eq:reduced model}
\end{equation}
where only four parameters $\theta_{reduced}=\left(x_{0,}x_{c},P_{x},\nu_{x}\right)$
will be inferred. We refer to this model as a ``reduced model''
from now on. Using less number of parameters decreases the iterations
required to reach equilibrium, as well as the computation time.

From the reduced model, the betatron tune is systematic higher than
that from the original model, due to the term $A(j)$ in the original
model shifts the sinusoidal frequency. The average tune difference
of all BPMs is as small as the $1.7\times10^{-6}$, which is sufficiently
small for almost any real application. inferred optics and its uncertainty
between the two models. What distinguishes the reduced model and the
original model, is the confidence of the inferred optics functions.

\begin{figure}
\includegraphics[width=1\columnwidth]{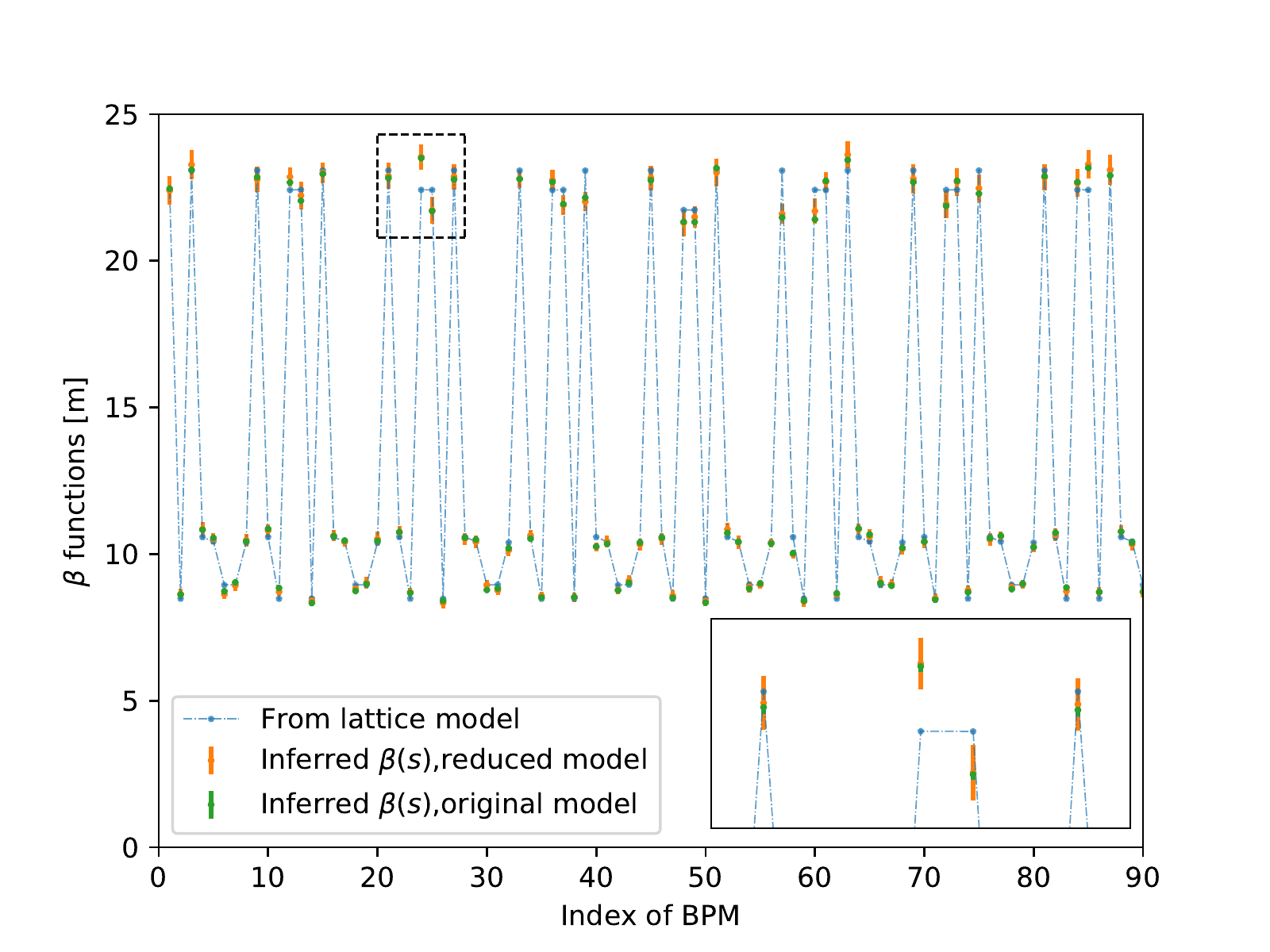}

\caption{The inferred beta function at first 90 BPMs with the original and
reduced model. \label{fig:inferred beta function compare}}
\end{figure}

Figure \ref{fig:inferred beta function compare} shows the comparison
of the inferred beta function of the reduced model and the original
model. With the reduced model (Eq. \ref{eq:reduced model}) we have
overlapped the mean value of the beta function from the mean values
of the original model, with about a 0.4\% standard deviation. However,
the uncertainty of the beta function on average increases by about
4.4 times compared with the original model. The betatron phase advance
of the reduced model also has larger uncertainties, ranging from 0.01
to 0.015 rad. They are about 5 times larger than that of the original
model. The uncertainty of the optics function may play an important
role in the lattice correction as pointed out in Ref \citep{Bayesian_linear_optics}.

It is worthwhile to note that the ``reduced model'' is not an incorrect
model. The results from the reduce model just have larger uncertainties.
They can be used as the prior distribution if we extend the model
to include the decoherence effect as in Eq. \ref{eq:model}.

\begin{figure}
\includegraphics[width=1\columnwidth]{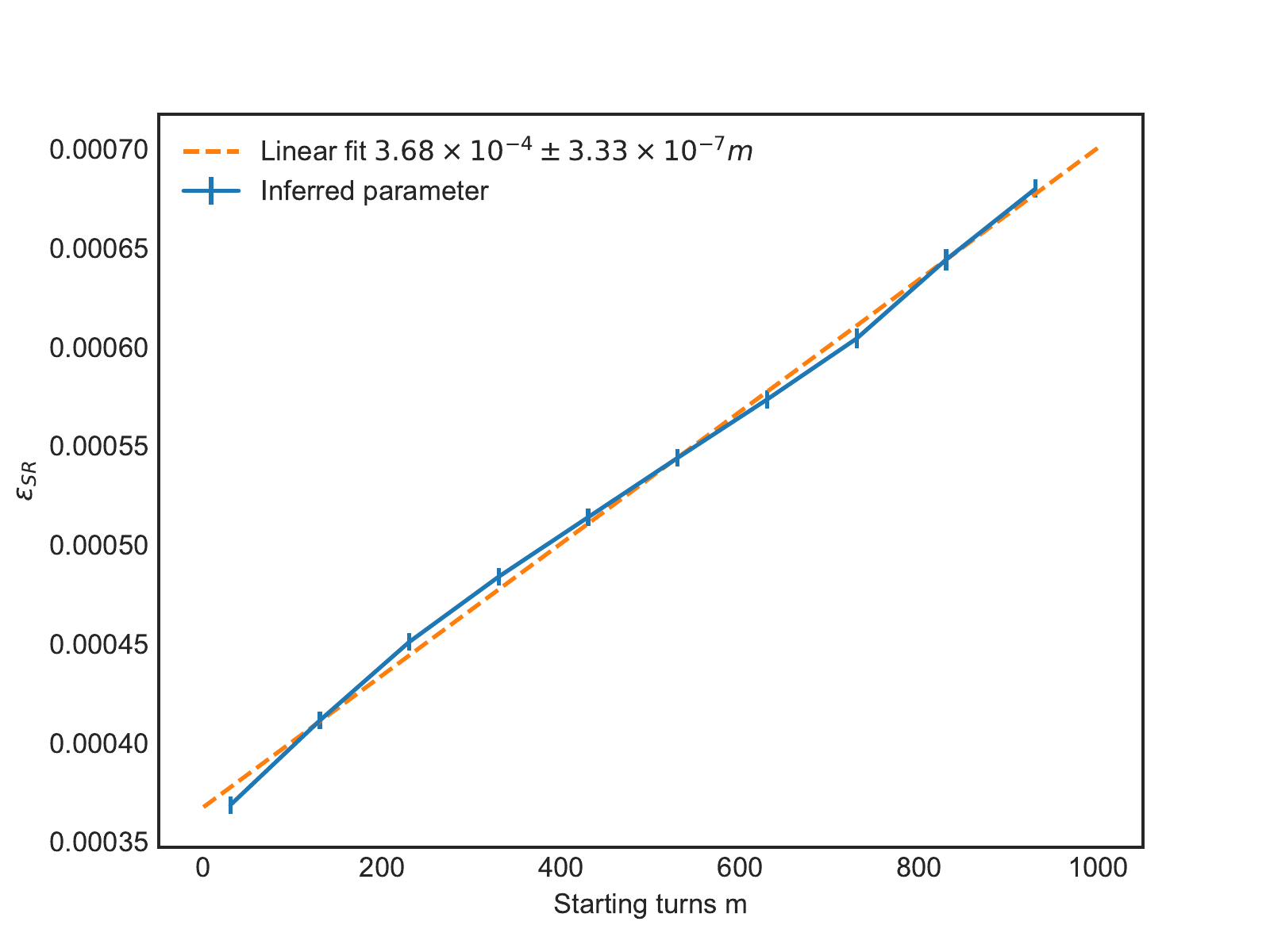}

\caption{The inferred $\epsilon_{SR}$ from 1000 turns data with varying starting
turns. \label{fig:inferred eps sr starting turn}}
\end{figure}

For the factor $A(j)$, it is also worthwhile to note that the choice
of the quadrature form $\exp\left(-\epsilon_{SR}j-\epsilon_{NL}^{2}j^{2}\right)$
is not arbitrary. On the one hand, we understand that these two terms
have their physics meaning. On the other hand, we may also get hint
only from the BPM data pretending to have no accelerator physics knowledge
on nonlinear decoherence. If $A(j)$ only has the form of the exponential
decay term:

\[
A(j)=A(0)\exp\left(-\epsilon_{SR}j\right)
\]
where $A(0)=1$. The envelope decay would be ``memoryless'', viz.
we expect the same parameter $\epsilon_{SR}$, no matter from which
turns we start our analysis. If we start our analysis from $m^{\text{th}}$
turn:
\begin{align*}
A(j+m) & =A(0)\exp\left(-\epsilon_{SR}\left(j+m\right)\right)\\
 & =A(0)\exp\left(-\epsilon_{SR}m\right)\exp\left(-\epsilon_{SR}j\right)\\
 & =A(m)\exp\left(-\epsilon_{SR}j\right)
\end{align*}
We can test by selecting 1000-turn data out of the total 2000 turns
with varying starting turns and used to infer the model which contains
$A(j)=\exp\left(-\epsilon_{SR}j\right)$ and found that the inferred
$\epsilon_{SR}$ is not a constant, as shown in Figure \ref{fig:inferred eps sr starting turn}.

\begin{figure*}
\includegraphics[width=0.33\textwidth]{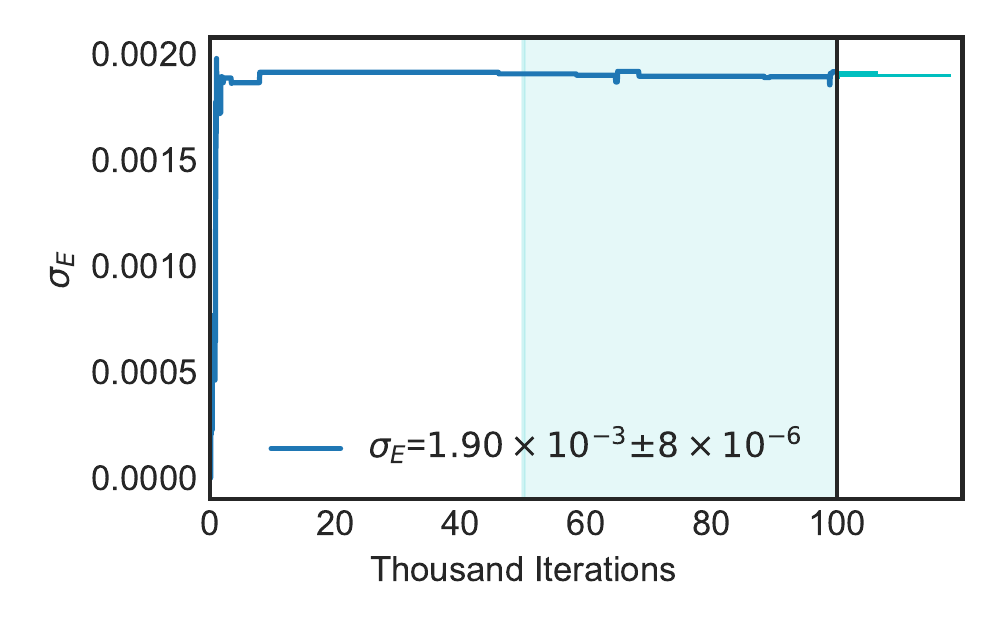}\includegraphics[width=0.33\textwidth]{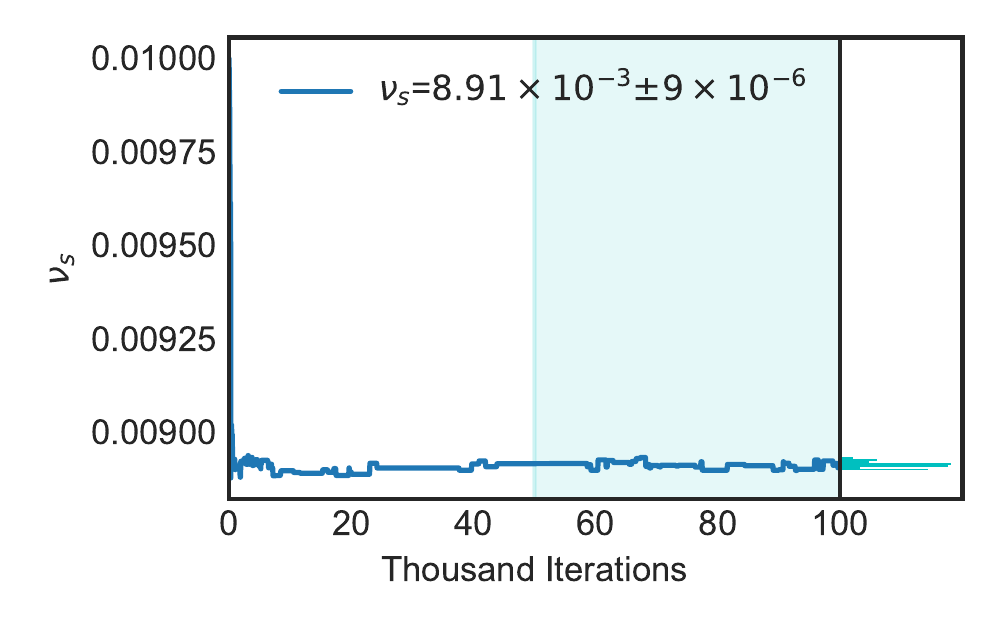}\includegraphics[width=0.33\textwidth]{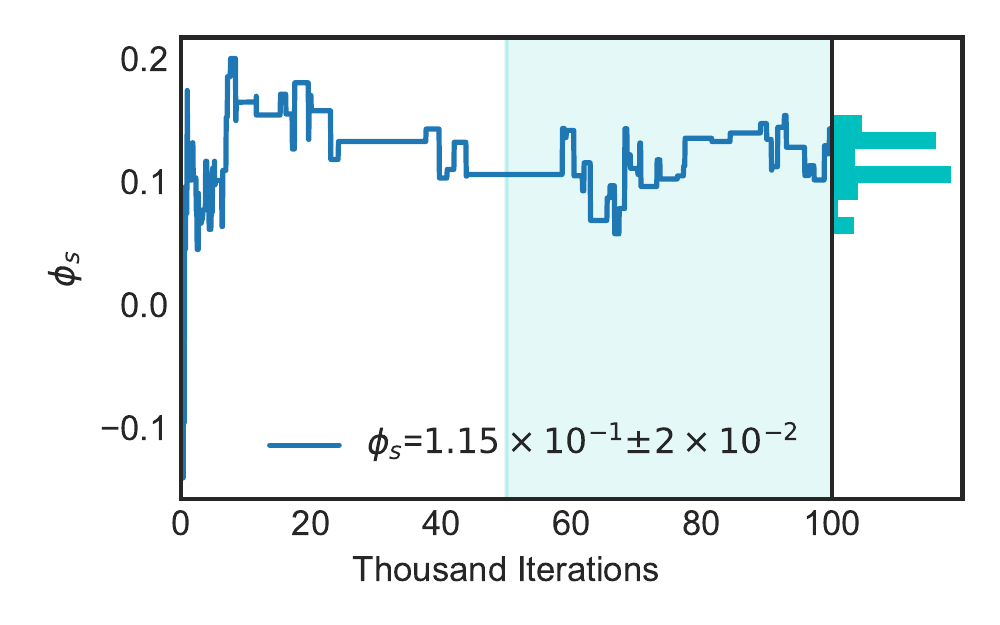}

\caption{The inference of the synchrotron oscillation parameters Left: $\epsilon_{E}=\xi\sigma_{E}$,
middle: $\nu_{s}$ right: synchrotron phase $\phi_{s}$.\label{fig:synchrobeta_model}}
\end{figure*}

Instead of a flat dependence on the starting turn $m$, the inferred
$\epsilon_{SR}$ is approximately a linear function of $m$ with a
significant slope. Therefore it suggests that the model is not ``memoryless''
and can not be modeled simply by an exponential decay function. We
may improve the model by adding an unknown function $f(j)$:
\begin{equation}
A(j)=A(0)\exp\left(-\epsilon_{SR}j-f(j)\right)\label{eq:guessed relation}
\end{equation}
If we start inference at turn $m$, factor $A(j)$ should reads:
\begin{equation}
A(j+m)=A(0)\exp\left(-\epsilon_{SR}\left(j+m\right)-f(j+m)\right)\label{eq:guessed relation starting m}
\end{equation}
Meanwhile, if the slope in Figure \ref{fig:inferred eps sr starting turn}
denoted as $\kappa$, the exponential decay rate starting at turn
$m$ is approximately $\epsilon_{SR}(m)=\epsilon_{SR}+\kappa m$.
$A(j+m)$ can be rewritten as:\begin{widetext}
\begin{align}
A(j+m) & =A(m)\exp\left(-\left(\epsilon_{SR}+\kappa m\right)j-f(j)\right)\nonumber \\
 & =A(0)\exp\left(-\left(\epsilon_{SR}+\kappa m\right)j-\epsilon_{SR}m-f(m)-f(j)\right)\nonumber \\
 & =A(0)\exp\left(-\epsilon_{SR}\left(j+m\right)-f(m)-f(j)-\kappa mj\right)\label{eq: a from observation}
\end{align}
\end{widetext}From Eq. \ref{eq:guessed relation starting m} and
\ref{eq: a from observation}, the extra term $f$ should satisfy:
\begin{equation}
f(j+m)=f(j)+f(m)+\kappa mj\label{eq:feed_down_f}
\end{equation}
in order to represent the linear slope in Figure \ref{fig:inferred eps sr starting turn}.
The simplest form of function $f$ is the quadrature form $f(j)\propto j^{2}$.
Therefore, learning from the data, we confirm that the simplified
model of decoherence factor $A\left(j\right)$ in Eq.\ref{eq:Damping factor form}
is a reasonable and simple choice.

Our model (as in Eq.\ref{eq:model}) certainly will not reflect all
physics hidden in the data. Figure \ref{fig:difference-dft} indicates
that there are still signals in the difference between the data and
the model. It is easy to identify the synchrotron sidebands and the
transverse coupling as stated in the previous sections. If this information
is needed, a more precise model can be adopted. For instance, we can
extend the model to extract the chromatic decoherence, by modifying
the decoherence term $A(j)$ as:

\[
A\left(j\right)=\exp\left(-\epsilon_{SR}j-\epsilon_{NL}^{2}j^{2}-\frac{\alpha^{2}(j)-\alpha^{2}(0)}{2}\right)
\]
with the new function $\alpha(j)$ has the form:
\[
\alpha(j)=\frac{2\xi_{x}\sigma_{E}}{\nu_{s}}\sin\left(\pi\nu_{s}j+\phi_{s}\right)
\]
which is representing the chromaticity decoherence \citep{Meller_decoherence}.
$\xi_{x}$ is the linear chromaticity of $x$-direction, $\sigma_{E}$
is the RMS energy spread, $\nu_{s}$ is the synchrotron tune and $\phi_{s}$
is the synchrotron phase. Using this synchro-betatron model, three
more parameters can be extracted, which are $\epsilon_{E}=\xi_{x}\sigma_{E}$,
$\nu_{s}$ and $\phi_{s}$, in additional to the parameters $\theta=\left(x_{0,}x_{c},P_{x},\nu_{x},\epsilon_{SR},\epsilon_{NL}\right)$
in the original model. Based on the fact that the synchro-betatron
mode is a small perturbation of the betatron motion, the mean values
of the inference results from the original model can be used as the
prior information to infer the synchro-betatron parameters. Figure
\ref{fig:synchrobeta_model} demonstrates the inference results of
the 9$^{\text{th }}$ BPM. The product $\xi_{x}\delta_{E}$ saturates
at $1.9\times10^{-3}$ with standard deviation of $8\times10^{-6}$.
This is consistent with the estimated chromaticity $\sim2$ and RMS
energy spread of $\sim9\times10^{-4}$. The inferred synchrotron tune
also meets the values from the lattice model. The minimization of
the model and the data is a weaker function of the synchrotron phase,
therefore the synchrotron phase has a large standard deviation from
its inferred value.

\begin{figure}
\includegraphics[width=1\columnwidth]{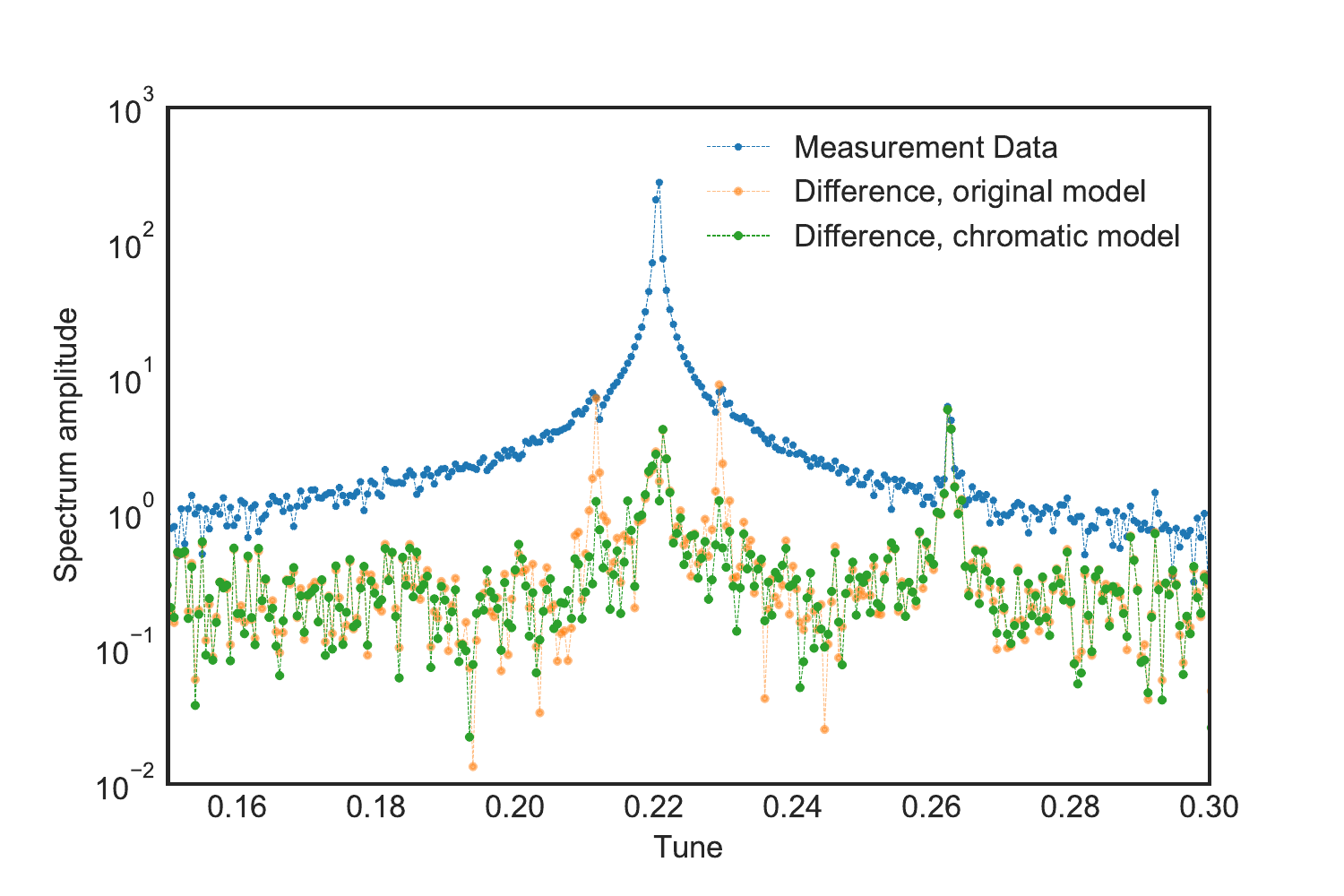}

\caption{Frequency spectrum of the measurement data (blue dots) and residue
between measurement and model with inferred parameters of the 9$^{\text{th }}$
BPM, using the original model (orange dots) and the model with chromatic
decoherence (green dots) \label{fig:difference-dft-synchro-beta}}
\end{figure}

Figure \ref{fig:difference-dft-synchro-beta} demonstrates that after
using the chromatic decoherence term in $A(j)$, we can eliminate
the two synchrotron sideband from the frequency spectrum of the difference
of the measurement and the model.

However, the convergence of inferring chromatic decoherence varies
among all BPMs. For some BPMs, very long iterations (larger than 200
thousand iterations) is needed, inferred synchrotron tuning from some
BPM may yield twice of the synchrotron tunes. This issue is likely
related to the fact that the chromatic decoherence is a much weaker
signal compared with the betatron motions. Actually, it is not wise
to extract very small signal using the Bayesian Inference. If the
synchrotron motion is of interest, we should design another experiment
with perturbed longitudinal motions and infer the unknowns with the
corresponding models. The statistics property obtained from this article
can be used as prior information of the transverse optics in the next
inference process. The Bayes's theorem enables us to sequentially
update our understanding of an accelerator by serials of measurements
and the evolution of the models. We will demonstrate this benefit
in future works.

\section*{Summary}

In this paper, we introduce a new approach to infer the parameters
of the accelerator model, using the measured TbT data in MCMC algorithm
based Bayesian Inference. It requires our prior knowledge, which includes
a proper accelerator model with parameters to be determined and our
belief (prior distribution) on the parameters. The MCMC algorithm
used in Bayesian Inference can generate samples from the posterior
distribution of the unknown parameters in the model using only a single
snapshot of measurement data. From the samples, the most probable
values of the parameters and their standard deviation are obtained.
Using the Bayesian approach, multiple data requisition is not necessary.
Therefore, pollution due to the slow drift of machine parameters and
the environment is mitigated or significantly reduced. In addition,
the model used in the inference can be easily extended or refined,
either from the accelerator physics knowledge or from the data analysis.
If a model needs to be extended from its original form, the posterior
distribution of the original model can serve as the prior distribution
of the extended one.

A proof-of-concept example has been demonstrated by exploring the
inference of optics functions of the betatron motion in the horizontal
plane, using measurement data of the NSLS-II electron storage ring.
There is no foreseen difficulty to apply the method to hadron rings.
A direct application of the reconstructed optics functions is the
optics correction. As pointed out in Ref \citep{Bayesian_linear_optics},
the statistical properties of the optics function will gain more insight
into the optics correction process by avoiding over-fitting. This
approach could also be potentially extended to analyzing more complicated
dynamics in accelerators, such as determining coupling coefficient,
analyzing the nonlinear dynamics properties, or estimate the reliability
of each instrumentation devices.
\begin{acknowledgments}
Work supported by the National Science Foundation under Cooperative
Agreement PHY-1102511, the State of Michigan and Michigan State University.
This research also used resources of the National Synchrotron Light
Source II, a U.S. Department of Energy (DOE) Office of Science User
Facility operated for the DOE Office of Science by Brookhaven National
Laboratory under Contract No. DE-SC0012704.
\end{acknowledgments}

\section*{Appendix A: Markov Chain Monte-Carlo Method}

Markov Chain Monte-Carlo (MCMC) is a powerful method, which can sample
the posterior distribution of the parameters of interest. It is especially
useful when in Bayesian Inference case (Eq.\ref{eq:bayes' theorem}),
since the marginal distribution $P\left(M\right)$ is usually impossible
to be directly calculated. The MCMC method constructs a Markov chain,
whose equilibrium distribution is proportional to the product of the
likelihood and prior distribution, and sample it using Monte-Carlo
method. The details of MCMC can be found in many text books, such
as Chapter 10 in Ref \citep{simulation_ross}. Here we only outline
the MCMC algorithm used this article.

We use the Hastings-Metropolis algorithm, an MCMC algorithm, to get
the sample of random variable $\theta^{i}$, with $i$ as the iteration
index, whose limiting probability is the posterior distribution $P\left(\theta\mid M\right)$.
The algorithm is detailed as below procedures:
\begin{enumerate}
\item Choose initial condition ($0^{\text{th}}$ iteration) $\theta^{0}$.
The choice does not affect the inference result. However, a reasonable
guess of the initial condition reduces the required iteration to reach
equilibrium.
\item Evaluate the $\pi(i)=P\left(M\mid\theta^{i}\right)\cdot P\left(\theta^{i}\right)$
for the $i^{\text{th}}$ iteration
\item Make Gaussian random walk centered at the value of $\theta^{i}$,
with preset step size as the standard deviation, to get the new trial
parameters $\theta^{t}$
\item Evaluate the $\pi(t)=P\left(M\mid\theta^{t}\right)\cdot P\left(\theta^{t}\right)$
\item Get a sample $u$ from uniform random distribution $[0,1]$
\item If $u<\min\left(\frac{\pi\left(t\right)}{\pi\left(i\right)},1\right)$,
the random walk is accepted, $\theta^{i+1}=\theta^{t}$; otherwise
$\theta^{i+1}=\theta^{i}$
\end{enumerate}
We continue this procedure for $N$ iterations. If after first $n\left(<N\right)$
iterations, the chain reaches its equilibrium, then the sequence $\left(\theta^{n+1},\theta^{n+2},\cdots,\theta^{N}\right)$
is the desired sampling of the posterior distribution of $\theta$.

\section*{Appendix B: Model for Nonlinear Decoherence}

\begin{figure}
\includegraphics[width=1\columnwidth]{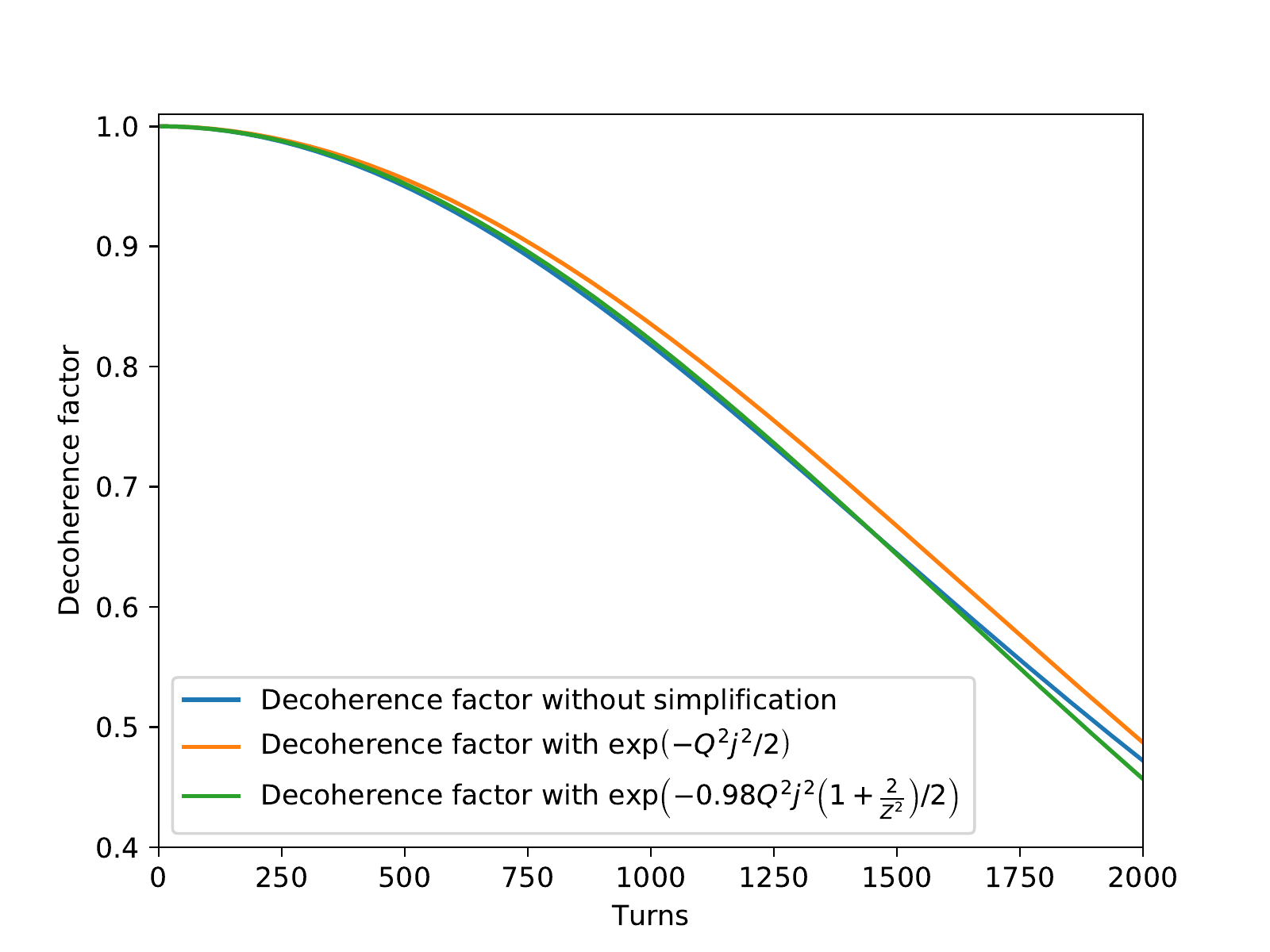}

\caption{The comparison of different decoherence model. \label{fig:comparision of decoherence model}}
\end{figure}

In Ref \citep{Meller_decoherence}, the decoherence factor of the
beam centroid motion due to nonlinearity is given by Eq. 20, which
can be written using the notation in this article:
\begin{equation}
A(j)=\frac{1}{1+Q^{2}j^{2}/Z^{2}}\exp\left[-\frac{Q^{2}j^{2}/2}{1+Q^{2}j^{2}/Z^{2}}\right]\label{eq:general_decoherence}
\end{equation}
where $Z=\beta\Delta x^{\prime}/\sigma_{x}$ denotes the the normalized
strength of the kick angle, $Q=-4\pi Z\nicefrac{d\nu}{da^{2}}$ represents
the effect of detuning and $a=\sqrt{\beta\epsilon}/\sigma_{x}$ is
the betatron amplitude scaled by rms beam size.

The Gaussian decoherence approximation is given by the Eq. 18, 
\begin{equation}
A(j)\approx\exp\left[-\frac{Q^{2}j^{2}}{2}\right]\label{eq:decoherence_gaussian_normal}
\end{equation}
 if the amplitude tune dependence is small and the observation time
is short. The condition for a such approximation is 
\begin{equation}
Z\gg QN\text{ or }4\pi\left|\frac{d\nu}{da^{2}}\right|N\ll1\label{eq:condition_gaussian_approximation}
\end{equation}
for $N$-turns data set.

The gaussian form of the decoherence factor is preferred in the Bayesian
inference since the only one parameter $Q$ is used, while in the
general form (Eq. \ref{eq:general_decoherence}), an additional parameter
$Z$ is needed.

In the NSLS-II example, the transverse kick strength $Z$, which excites
the betatron oscillation, is about 3.7. The NSLS-II model predict
that the amplitude tune dependence $\nicefrac{d\nu}{da^{2}}$ is $-1.28\times10^{-5}$
by assuming the horizontal emittance as $2.1\times10^{-9}$ nm rad,
hence $QN/Z$ equals 0.32 for 2000 turns data, which is not much less
than 1. Therefore, the condition (Eq. \ref{eq:condition_gaussian_approximation})
is not fully satisfied. The difference between two forms (Eq. \ref{eq:general_decoherence}
and \ref{eq:decoherence_gaussian_normal}) is shown from the blue
and orange lines in Figure\ref{fig:comparision of decoherence model}.Since
the difference is small over the 2000 turns, we are motivated to seek
for an alternated Gaussian form by modifying the exponential term.

Slightly modifying Eq. \ref{eq:decoherence_gaussian_normal}, we use
the following Gaussian form 
\begin{equation}
A(j)\approx\exp\left[-\frac{rQ^{2}j^{2}}{2}(1+\frac{2}{Z^{2}})\right]\label{eq:decoherence_gaussian_modified}
\end{equation}
which deviates from the general decoherence form (Eq. \ref{eq:general_decoherence})
by $\mathcal{O\text{\ensuremath{\left(Q^{4}j^{4}\right)}}}$ when
the factor $r=1$. The green line in Figure\ref{fig:comparision of decoherence model}
shows that by adjusting the factor $r$, eg. $r=0.98$ in the plot,
we can achieve approximate the general decoherence factor within the
length of dataset.

\bibliographystyle{apsrev4-1}
\bibliography{ref}

\end{document}